\documentclass[10pt,letterpaper]{article}
\usepackage{opex3b}
\usepackage{geometry}
\usepackage{color, graphicx}
\usepackage{mathptmx, courier, helvet}
\usepackage{amsmath}
\usepackage{subfigure}
\usepackage{algorithm2e}

\newcommand{\mycitet}[1]{\cite{#1}}
\newcommand{\mycitep}[1]{\cite{#1}}
\newcommand{\myeq}[1]{Eq. (\ref{#1})}
\newcommand{\myfig}[1]{Fig. \ref{#1}}

\newcommand{\intd}[1]{\ensuremath{\,\mathrm{d}#1}}

\begin{document}
\title{Boundary diffraction wave integrals for diffraction modeling of external occulters}
\author{Eric Cady}
\address{Jet Propulsion Laboratory, California Institute of Technology, 4800 Oak Grove Drive, \\ Pasadena, CA, 91109 USA}
\email{eric.j.cady@jpl.nasa.gov}

\begin{abstract}
An occulter is a large diffracting screen which may be flown in conjunction with a telescope to image extrasolar planets.  The edge is shaped to minimize the diffracted light in a region beyond the occulter, and a telescope may be placed in this dark shadow to view an extrasolar system with the starlight removed.   Errors in position, orientation, and shape of the occulter will diffract additional light into this region, and a challenge of modeling an occulter system is to accurately and quickly model these effects.  We present a fast method for the calculation of electric fields following an occulter, based on the concept of the boundary diffraction wave: the 2D structure of the occulter is reduced to a 1D edge integral which directly incorporates the occulter shape, and which can be easily adjusted to include changes in occulter position and shape, as well as the effects of sources---such as exoplanets---which arrive off-axis to the occulter.  The structure of a typical implementation of the algorithm is included.
\end{abstract}


\section{Introduction}

One of the major goals of the coming decades is to directly image a terrestrial exoplanet.  Direct imaging allows spectral information about the planet's atmosphere to be extracted from the spectrum of reflected light, including the potential presence of biomarkers, gases whose presence and abundance may indicate the presence of life.  Detecting planets directly is a difficult task, however, for two reasons: intensity ratio (\emph{contrast}) and angular resolution.  An Earth-twin which orbits a Sun-twin at 1AU emits $10^{10}$ times less flux than the parent star in the visible band; if the system was located at 10 parsecs from Earth, the angular separation of the two objects would be 100mas, which for most proposed space telescopes puts the separation under $4 \lambda/D$.  Specialized methods are required to remove this flux at these small working angles.

One proposed method of suppressing the starlight is an \emph{occulter}, a spacecraft with a shaped edge flown in front of the telescope to block the starlight before it arrives at the telescope.  The occulter size (tens of meters) and distance (tens of thousands of kilometers) are chosen so the angular extent of the occulter is smaller than some desired angle, on the order of 100 milliarcseconds, so exoplanets in orbit about the star will still be visible.   The edge is shaped to control diffraction, with the form chosen to suppress the light to a factor of $10^{10}$ across the telescope aperture and over a wide spectral passband.

A major advantage of this approach, compared to an internal coronagraph, is that it significantly loosens the tolerances on the optical quality and thermal stability of the telescope optics, as well as eliminating the need for a wavefront control system for active correction.   Conversely, the occulter itself must be built, deployed, and flown in formation to its own set of tolerances, and demonstrating this capability is an area of active research \mycitet{Kas11}.  These tolerances include limits on permanent errors in the occulter shape due to manufacturing and deployment errors; transient shape deformations from thermal fluctuations and vibration; and position and orientation changes of the occulter relative to the telescope-target axis from formation flying \mycitep{Sha10, Gla10}.

The scale of an occulter---and the distance it must maintain with respect to its target---precludes optical testing on the ground. All of these effects must be modeled in order to build an error budget and verify that an occulter can satisfy these requirements.  The capture of the large dynamic range in the resulting field---$10^{5}$ or greater in amplitude---drives the accuracy of the propagator used for the models, while the modeling of time-varying thermal shape deformations and closed-loop formation flying in broadband light drive the speed of the calculation.  Some of the most in-depth simulated observations can take from hours to days to run, and the ability to quickly and accurately model propagation past an occulter is thus a major enabling factor in the characterization of occulter system performance.  Section \ref{sec:current} gives an overview of the current techniques; the new technique presented in this work is described in Sections \ref{sec:bdw} and \ref{sec:vecpot}, and a suggested implementation scheme, along with a pseudocode overview of the algorithm and some computational results, are given in Section \ref{sec:imp}.

\section{Current occulter modeling techniques} \label{sec:current}

\begin{figure}[ht]
\begin{center}
\includegraphics[width=0.8\columnwidth]{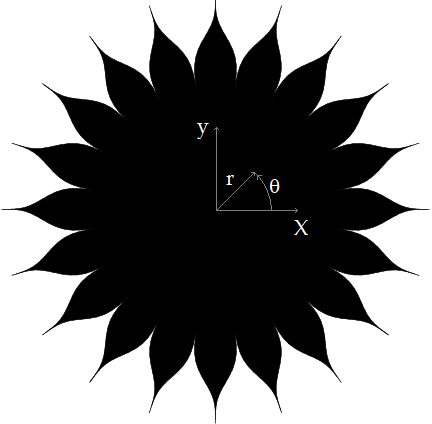}
\end{center}
\caption{An example of a typical occulter with 20 petals, with the associated Cartesian and polar coordinate systems.} \label{fig:binocc}
\end{figure}

Occulter designs take the general form of a solid central disk surrounded by $N_{p}$ identical tapering structures (\emph{petals}) around the edge, giving the whole structure the general appearance of a flower.  (An example is shown in \myfig{fig:binocc}.)  This shape is a result of their provenance as (0,1)-valued approximations to apodizers; the general method to design an occulter is to determine a smooth apodization profile $A(r)$ which can provide the necessary starlight suppression, and then convert it into a binary shape with a sufficient number of petals so that performance is not undermined.  The actual form of $A(r)$ can be determined by optimization \mycitet{Van07, Cad08} or selecting an existing functional form \mycitet{Cop00, Cas06}.

For an occulter designed in this manner, the region $\Omega$ in which the occulter is opaque can be written directly:
\begin{align}
    \Omega &= \{(r, \theta): 0 \leq r \leq R, \theta \in \Theta(r)\} \\
    \mathrm{where}\quad\Theta(r) &= \bigcup^{N_{p}-1}_{n=0}\left[\frac{2 \pi n}{N_{p}} - \frac{\pi}{N_{p}}A(r), \frac{2 \pi n}{N_{p}} + \frac{\pi}{N_{p}}A(r)\right], \label{td3}
\end{align}
with $(r, \theta)$ being polar coordinates in the plane of the occulter.  Each of the $[\ldots,\ldots]$ is a set of points which defines the outline of a single petal; $\bigcup$ denotes the union of these sets, which sets the boundary for the entire occulter.  (We denote this boundary as $\partial \Omega$, as it will be used later.)

For a typical occulter, the occulter-telescope distance $z = 5\times10^{7}$m; the wavelength $\lambda = 5\times10^{-7}$m; the occulter radius $R = 25$m; and the maximum excursions in the plane of the telescope $|\xi|$ and $|\eta| \leq 3$m.  These values place the occulter well within the Fresnel regime; that is, the region in which it is valid to approximate the exponent in the propagation integral with the first two terms of a power series (the Fresnel approximation).  This approximation is suitable when \mycitep{Goo96}.
\begin{equation}
z^3 \gg \frac{\pi}{4 \lambda}[(x - \xi)^2 + (y - \eta)^2]^2_{\mathrm{max}}.
\end{equation}

Further, a standard propagation integral computes the electric field past a finite aperture; the occulter, on the other hand, is a finite obstruction with an effectively infinite aperture.  Here, we can turn to Babinet's theorem \mycitep{Bor99}, which notes that field propagated through an aperture and the field propagated past its complement sum to the field that would have resulted were no obstruction present at all.

If a plane wave with amplitude $A$ and wavelength $\lambda$ is normally-incident on the occulter, as would be the case for an occulter properly aligned with a target star, the electric field at a distance $z$ past the occulter would be written with the Fresnel approximation and Babinet's principle as
\begin{align} \label{2Deq}
    U(\xi, \eta) &= A \exp{\left(\frac{2 \pi i z}{\lambda}\right)}\left(1-\frac{1}{i \lambda z} \int\int_{\Omega} \exp{\left\{\frac{\pi i}{\lambda
    z}\left[(r \cos{\theta}-\xi)^2 + (r \sin{\theta}-\eta)^2\right]\right\}} r \intd{r} \intd{\theta}\right) \nonumber\\
    &= A \exp{\left(\frac{2 \pi i z}{\lambda}\right)}\left(1-\frac{1}{i \lambda z} \int\int_{\Omega} \exp{\left\{\frac{\pi i}{\lambda
    z}\left[(x-\xi)^2 + (y-\eta)^2\right]\right\}} \intd{x} \intd{y}\right),
\end{align}
where $\xi$ and $\eta$ are Cartesian coordinates at the downstream location. The ``$1 -$'' term before the integral results from the use of Babinet's principle. Errors in (for example) occulter shape will modify $\Omega$, but \myeq{2Deq} will hold regardless. Computing the integral for points in the plane of a telescope aperture is the key step in modeling the performance of an occulter.

One could attempt this integral by brute force:  discretizing the shape of the occulter over a sufficiently fine grid to capture all of the necessary structure, and propagating this to the telescope aperture with direct integration or a matrix Fourier transform\mycitep{Sou07}.  This approach is not an easy task, as the occulter will be tens of meters across, with tolerances at the level of tens of microns\mycitep{Sha10}, requiring a grid up to $10^{6} \times 10^{6}$ or more to capture the full shape of the profile, not including array padding.

A better approach is to take advantage of the structure of the occulter: in the absence of errors, it has an $N_{p}$-fold symmetry in the azimuthal direction, and most errors introduce small perturbations to that.  Taking advantage of the symmetry leads to three main approaches to modeling propagation past an external occulter:
\begin{enumerate}
\item Integrals in $r$: The Bessel function expansion \mycitep{Van07} does the integral over $\theta$ explicitly, reducing the 2D integral to a series of single integrals in $r$ whose contributions fall off exponentially fast near the optical axis, which makes computation extremely quick in this region.  The series can be modified to incorporate some shape errors \mycitet{Cad10b} at the expense of slower convergence.  Unfortunately, it is slower at locations far from the optical axis, and requires a good deal of effort to incorporate many types of errors, such as occulter tilt, into the model.  Aside from the propagation modeling, the first, radially-independent term in this series is used in optimization approaches \mycitep{Van07, Cad08} to design occulter shapes.
\item Integrals in $\theta$: The Dubra-Ferrari integral \mycitet{Dub99} starts one step back from the Fresnel form of \myeq{2Deq}, at the Raleigh-Sommerfeld diffraction integral, and evaluates the integral over $r$ to produce a pair of single integrals in angular coordinates, one of which holds for points inside the geometric extent of the starshade, and one which holds outside.  This approach can incorporate most errors straightforwardly and spends the same amount of time to determine the field any point in the telescope aperture plane.  Because of the manner in which the integrals are segmented, however, the integrands are more difficult to determine correctly for points physically outside the extent of the starshade.
\item Perturbations to a nominal shape: The slit approach \mycitet{Dum09} assumes the electric field for an occulter with no errors has been calculated by one of the above methods, and approximates the difference between the shape with and without errors as a series of long, thin boxes whose Fresnel transforms can be calculated quickly and exactly.  This method is very fast but necessarily approximate.
\end{enumerate}

In this paper, we propose a fourth method---the boundary-diffraction-wave approach---which seems to out-perform the three extant methods for general purpose use.  It maintains the constant-time property of the Dubra-Ferrari integral with a 2-3$\times$ improvement in computational speed; see Sec. \ref{subsec:comp} for further discussion.  It incorporates all errors on the shape of the occulter natively, and can be written straightforwardly to include position and orientation errors, as well as model waves from off-axis sources, such as an exoplanetary system or the finite diameter of the star.  It can also take advantage of the structure of the integral to evaluate the field at multiple wavelengths with little additional computation.  The approach of defining the occulter solely in terms of its boundary is a good fit for tolerancing the sorts of errors that are expected for an occulter.  Manufacturing errors can be applied by specifically deforming desired edge regions; errors in petal placement can be included by translating and rotating only the boundary points associated with that petal.

\section{The boundary diffraction wave formulation} \label{sec:bdw}

Most diffraction approaches to propagation through a finite aperture find their basis in the Huygens-Fresnel principle, which expresses the electric field at a point following the aperture as the sum over spherical waves emitting from every point within the aperture.  However, a different representation was introduced by Maggi and Rubinowicz which splits the field into two parts, a part which propagates according to geometrical optics and a part which incorporates the effect of diffraction from the boundary.  This boundary-diffraction-wave (BDW) representation was codified by the work of Miyamoto and Wolf \mycitep{Miy62, Miy62a}, who showed that the boundary integral could be represented for any incident field entirely by a line integral around a vector potential.  We will review their analysis briefly here.

Miyamoto and Wolf \mycitet{Miy62} begin by writing the Kirchhoff formulation (see \emph{e.g.} \mycitet{Bor99}) for the diffraction integral from the aperture.  Consider a volume of space, bounded by a surface $S$; the field at any point $P$ within the volume may be expressed as an integral over all surface points $Q$ on $S$:
\begin{equation}
    U(P) = \frac{1}{4 \pi}\iint_{S} \left(U(Q) \nabla_Q \frac{\exp{(i k s)}}{s} - \frac{\exp{(i k s)}}{s} \nabla_Q U(Q)\right) \cdot n \intd{S} \equiv \frac{1}{4 \pi}\iint_{S} \mathbf{V}(Q,P) \cdot n \intd{S}.
\end{equation}
Here we let $s$ be the distance between $P$ and $Q$, $n$ be a vector normal to $S$ at point $Q$, and $\nabla_Q$ is the gradient evaluated at point $Q$. In the case of the occulter, the volume is the space $z \geq 0$, and $S$ stretches over the $z = 0$ plane, as well as having a component at infinity which is assumed to vanish.  (Showing that this term vanishes requires some minor additional assumptions; see Sec. 8.3.2 of \mycitet{Bor99}).

\begin{figure}[ht]
\begin{center}
\includegraphics[width=1\columnwidth]{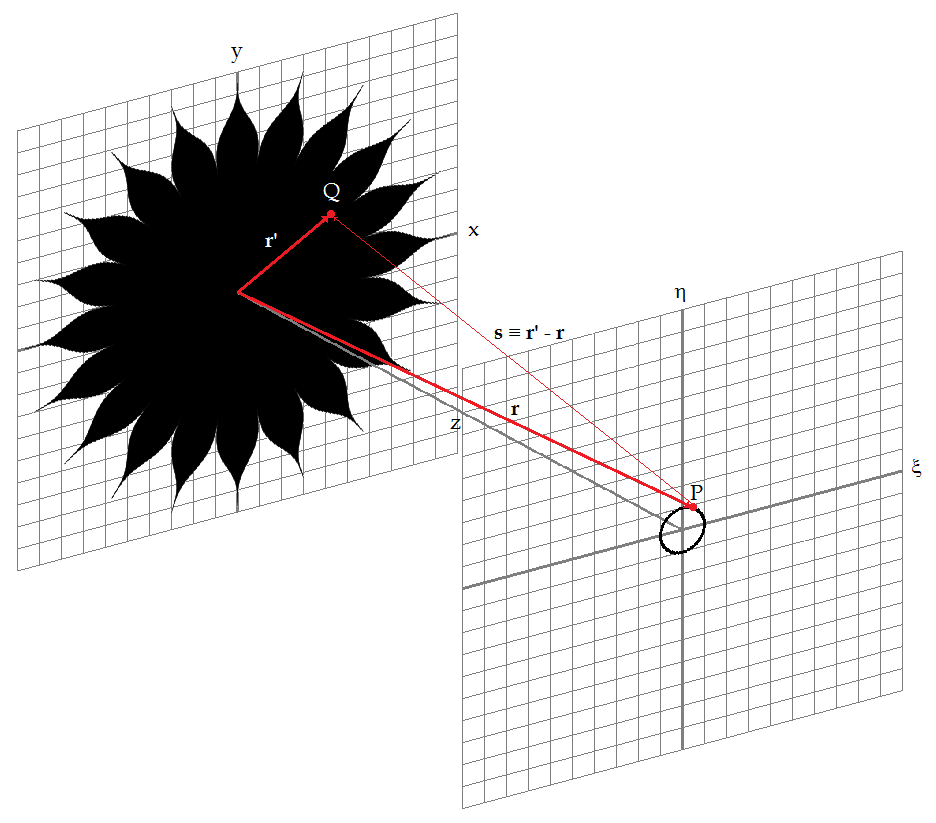}
\end{center}
\caption{A diagram of the points, coordinate systems, and vectors used in this paper.  The left grid is in the plane of the occulter, and the right grid in the plane of the telescope aperture.} \label{fig:coo}
\end{figure}

It can then be shown \mycitet{Miy62} that for $\mathbf{V}$ of this form, there exists a vector potential $\mathbf{W}$, such that $V = \nabla \times \mathbf{W}$, and for plane waves incident on an aperture, that associated vector potential is
\begin{align}
\mathbf{W}(Q, P) = \frac{1}{4\pi} A \exp{(i k \mathbf{p}\cdot\mathbf{r}')} \frac{\exp{(i k s)}}{s} \frac{\hat{s}\times\mathbf{p}}{1 + \hat{s}\cdot\mathbf{p}}, \label{vecPot} \\
U(P) = \iint_{S} \nabla \times \mathbf{W}(Q,P) \cdot n \intd{S}.  \label{Ueq}
\end{align}
Here $\mathbf{r}$ is the vector from the origin to $P$; $Q$ is a point within the aperture on $S$, and $\mathbf{r'}$ is the vector from the origin to $Q$.  $k = 2 \pi/\lambda$ is the wavenumber and $\lambda$ is the wavelength under consideration.  The various $s$-variables become
\begin{align}
\mathbf{s} &\equiv \mathbf{r'} - \mathbf{r}, \\
s &= ||\mathbf{s}||, \\
\hat{s} &= \frac{\mathbf{s}}{s}.
\end{align}
and we assume the form of the plane wave in the vector direction $\mathbf{p}$ to be
\begin{equation} \label{planeWave}
    U_0(P) = A \exp{(i k \mathbf{p}\cdot\mathbf{r})},
\end{equation}
with amplitude $A$.  A coordinate diagram is shown in \myfig{fig:coo}.

Next, we consider for the moment the case of an opaque screen with an opening over the region $\Omega$, a subregion of $S$ with boundary $\partial \Omega$.  We recall here that $S$ is the entire bounding surface of the half-plane $z \geq 0$, including the $z= 0$ plane; $\Omega$ is the section of it on which the occulter lies.  (We are not considering the occulter case at the moment, but its complement; we will return to the occulter shortly.)  In this case, the electric field from \myeq{Ueq} across the aperture is:
\begin{equation}
U_{\mathrm{ap}}(P) = \iint_{\Omega} \nabla \times \mathbf{W}(Q,P) \cdot n \intd{\Omega},
\end{equation}
Our approach is to reduce this integral to a line integral, using Stokes' Theorem.  To do this, we consider singularities of the vector potential.

\begin{figure}[ht]
\begin{center}
\includegraphics[width=1\columnwidth]{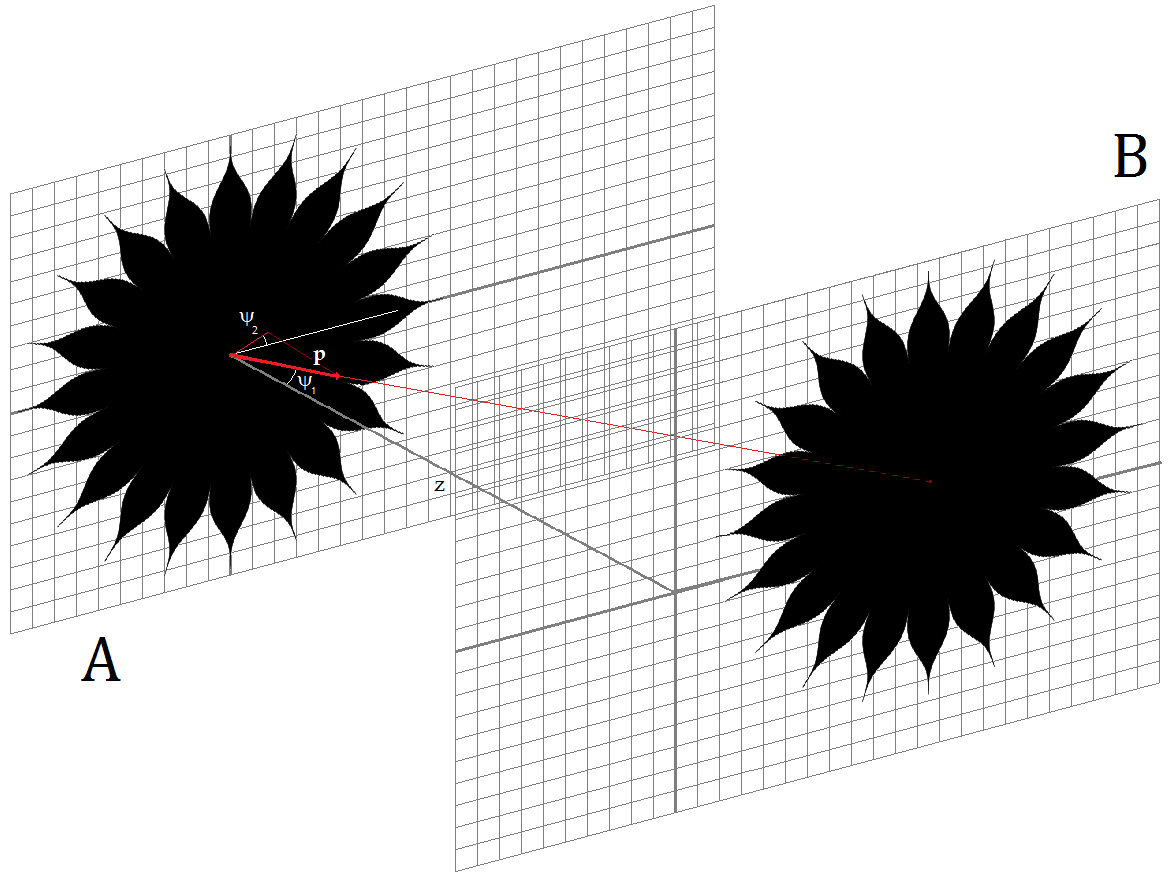}
\end{center}
\caption{A diagram showing the projected extent of the aperture given by geometric optics.  A plane wave incident on the aperture in plane A with vector direction $\mathbf{p}$ produces a shifted aperture in plane B.  Two angles which parameterize the direction of the plane wave, $\psi_{1}$ and $\psi_{2}$, are shown as well.} \label{fig:geom}
\end{figure}

In general, $\mathbf{W}$ will have singularities somewhere on $S$ for a given $P$, as otherwise the field in the half-space past the aperture will be zero.  In the specific case of the plane wave, the lone singularity occurs at $1 + \hat{s}\cdot \mathbf{p} = 0$, which occurs only when $\hat{s}$ is parallel to the propagation direction of the plane wave; thus, the singularity will fall in the aperture \emph{only} for points $P$ which fall in the beam as dictated by geometric optics.  (\myfig{fig:geom} shows an example of this.) Miyamoto and Wolf \mycitet{Miy62a} show that when Stokes' Theorem is applied, we get two distinct cases for the field at point $P$, depending on the singularity location:
\begin{align} \label{Udef}
U_{\mathrm{ap}}(P) &= A \exp{(i k \mathbf{p}\cdot\mathbf{r})} + U^{(B)}(P) \; \mathrm{for\;points\;inside\;the\;extent\;of\;the\;aperture} \\
 & = U^{(B)}(P) \; \mathrm{otherwise,} \nonumber
\end{align}
where $U^{(B)}(P)$ is the counterclockwise line integral about the edge of $\Omega$:
\begin{equation} \label{Ub}
U^{(B)}(P) = \int_{\partial \Omega} \mathbf{W}(Q,P) \cdot \ell \intd{\ell}.
\end{equation}
Here, $\ell$ is a unit vector in the direction tangent to the edge at any point, and $\intd{\ell}$ is a differential element of the boundary.

This formulation gives the field following an aperture, but in our case we have the opposite: the occulter is opaque, and the surroundings are open.  Here we return to Babinet's theorem.  As we started with the plane wave in \myeq{planeWave}, this implies our occulter field $U(P)$ satisfies
\begin{equation} \label{babinet}
    U_0(P) = U(P) + U_{\mathrm{ap}}(P),
\end{equation}
or
\begin{align} \label{fullU}
U(P) &= -\int_{\partial \Omega} \mathbf{W}(Q,P) \cdot \ell \intd{\ell} \;\; \mathrm{for\;points\;in\;the\;geometric\;shadow\;of\;the\;occulter,} \\
 & = A e^{i k \mathbf{p}\cdot\mathbf{r}} - \int_{\partial \Omega} \mathbf{W}(Q,P) \cdot \ell \intd{\ell} \; \mathrm{otherwise.} \nonumber
\end{align}
The part of \myeq{fullU} which depends on geometric optics is termed $U^{(G)}(P)$:
\begin{align} \label{1stUg}
U^{(G)}(P) &= 0 \; \mathrm{for\;points\;inside\;the\;extent\;of\;the\;aperture} \\
 & = U_0(P) \; \mathrm{otherwise.} \nonumber
\end{align}
This equation is the general form of the occulter field, in the boundary-diffraction-wave formulation.

\subsection{Vector potentials for occulter propagation} \label{sec:vecpot}

From this suitably general representation, we can make appropriate substitutions for the specific case of an occulter; these will simplify computation of electric fields later. We define the vector from the origin to $P$ as $\mathbf{r} = (\xi, \eta, z)$; $\mathbf{r'}$, the vector from the origin to $Q$, is defined as $\mathbf{r'} = (x, y, 0)$.  (See again \myfig{fig:coo}.)  We can write the $s$-variables explicitly as
\begin{align}
\mathbf{s} &\equiv \mathbf{r'} - \mathbf{r} =  (x - \xi, y - \eta, -z), \\
s &= ||\mathbf{s}|| = [z^2 + (x-\xi)^2 + (y- \eta)^2]^{1/2}, \label{sDef}\\
\hat{s} &= \frac{\mathbf{s}}{s}.
\end{align}

We consider first the case where the plane wave is normally-incident on the occulter, as starlight would be when the system is correctly aligned.  Given this, our vector terms can be written as
\begin{align}
\mathbf{p} &= (0, 0, 1), \\
\mathbf{p}\cdot \mathbf{r'} &= 0, \\
\hat{s} \times \mathbf{p} &= \frac{1}{s}(y-\eta, -(x-\xi), 0), \\
1 + \hat{s} \cdot \mathbf{p} &= 1 - \frac{z}{s},
\end{align}
and the potential in \myeq{vecPot} becomes
\begin{align}
\exp{(i k s)} &= \exp{(i k z)} \exp{[i k (s-z)]}, \\
\mathbf{W}(Q, P) &= \frac{1}{4\pi} A \exp{(i k z)} \frac{\exp{[i k (s-z)]}}{s-z} \hat{s}\times\mathbf{p}.  \label{W1}
\end{align}

More generally, an off-axis source at an angle of $\psi_1$ off-axis and $\psi_2$ from the x-axis will have
\begin{align}
\mathbf{p} &= (\sin{\psi_1} \cos{\psi_2}, \sin{\psi_1} \sin{\psi_2}, \cos{\psi_1}), \\
\mathbf{p}\cdot \mathbf{r'} &= x \sin{\psi_1} \cos{\psi_2} + y \sin{\psi_1} \sin{\psi_2}.
\end{align}
Before the expression for the vector potential becomes too complicated, we will define three intermediate variables ($f$, $g$, and $h$) which we can substitute into \myeq{sDef}; these choices will greatly simplify subsequent computations.
\begin{align}
f &= z\sin{\psi_1} + (x - \xi)\cos{\psi_2} \cos{\psi_1} + (y - \eta) \sin{\psi_2} \cos{\psi_1}, \label{fgh}\\
g &= -(x - \xi) \sin{\psi_2} + (y - \eta) \cos{\psi_2}, \\
h &= z \cos{\psi_1} - (x - \xi)\cos{\psi_2} \sin{\psi_1} - (y - \eta) \sin{\psi_2} \sin{\psi_1}  \equiv \mathbf{s} \cdot \mathbf{p}.
\end{align}
Substituting these variables gives:
\begin{align}
s &= [f^2 + g^2 + h^2]^{1/2}, \\
1 + \hat{s} \cdot \mathbf{p} &= 1 - \frac{h}{s}, \\
\hat{s} \times \mathbf{p} = \frac{1}{s} &\left(f \sin{\psi_2} + g\cos{\psi_2} \cos{\psi_1}, \right. \\
        & -f \cos{\psi_2} + g\sin{\psi_2} \cos{\psi_1},  \nonumber\\
        & \left.-g \sin{\psi_1}\right) \nonumber \\
\exp{(i k h)} &= \exp{[i k z \cos{\psi_1} - i k (x - \xi)\cos{\psi_2} \sin{\psi_1} - i k (y - \eta) \sin{\psi_2} \sin{\psi_1}]} \nonumber \\
&=\exp{(-i k \mathbf{p}\cdot\mathbf{r}')} \exp{(i k z \cos{\psi_1})} \exp{[i k (\xi \cos{\psi_2} + \eta \sin{\psi_2}) \sin{\psi_1}]}, \\
\mathbf{W}(Q, P) &= \frac{1}{4 \pi} A \exp{(i k z \cos{\psi_1})} \exp{[i k (\xi \cos{\psi_2} + \eta \sin{\psi_2}) \sin{\psi_1}]} \frac{\exp{[i k (s-h)]}}{s-h} \hat{s}\times\mathbf{p},
\end{align}
a similar relation to \myeq{W1}.  This equation is still exact; for $\psi_1 \ll 1$, the usual case for objects in an exoplanetary system, we can note that
\begin{align}
\frac{1}{s(s-h)} &= \frac{1}{f^2 + g^2}\left(1 + \frac{1}{[1 + \frac{f^2 + g^2}{h^2}]^{1/2}}\right) \approx \frac{2}{f^2 + g^2},\\
s - h &\approx \frac{f^2 + g^2}{2 h},
\end{align}
and so the potential becomes
\begin{align} \label{fullW}
\mathbf{W}(Q, P) &\approx \frac{1}{2\pi} A \exp{(i k z \cos{\psi_1})} \exp{[i k (\xi \cos{\psi_2} + \eta \sin{\psi_2}) \sin{\psi_1}]} \frac{\exp{\left(i k \frac{f^2 + g^2}{2 h}\right)}}{f^2 + g^2} \mathbf{v}(f,g), \\
\mathbf{v}(f,g) = &\left(f \sin{\psi_2} + g\cos{\psi_2} \cos{\psi_1}, \right. \\
        & -f \cos{\psi_2} + g\sin{\psi_2} \cos{\psi_1},  \nonumber \\
        & \left.-g \sin{\psi_1}\right) = \mathbf{s} \times \mathbf{p}. \nonumber
\end{align}
This form proves to be particularly useful, as $(f^2 + g^2)$ and $\mathbf{v} \cdot \ell$ can be calculated without loss of precision, as neither will be of order $z$ for the small $\psi_1$ case.  A typical occulter might have $z = 5\times 10^{7}$m, $R = 25$m, and $|\xi|$ and $|\eta| \leq 3$m, and $k \approx 10^{7}$m$^{-1}$; a representative exoplanet target for this occulter might have $\psi_1 = 5 \times 10^{-7}$ rad.  We note that exponential terms of order $kz$ have been separated from smaller terms; these should be calculated independently, as $kz \sim 10^{14}$, and combining terms will lose precision in evaluating the exponent.

\section{Efficient implementation} \label{sec:imp}

Evaluating the field at each downstream point requires the determination of two parts: the geometric field and the boundary field.  We assume, to begin, that we are given a set of $N$ points representing the edge of the occulter which form a simple closed curve, as well as a list of $M$ downstream points and a list of $L$ wavelengths at which the field is to be determined.  We also assume that the edge is sufficiently well-sampled that the section of the edge between explicitly-specified points is well-modeled by a linear segment, as this allows us to use the midpoint rule for numerically approximating the integral.  Given the slowly-changing shape over the majority of the petal, this is a good assumption; small regions such as petal tips may be specified more finely.

The geometric field is still a plane wave of form \myeq{planeWave} outside the aperture, and is $0$ everywhere inside; finding the geometric field thus becomes a problem of checking whether a downstream point falls behind it or not.  We do this by treating the edge as a polygon and finding the winding number of the downstream point $(\xi_0, \eta_0)$; this can be done efficiently with an $O(N)$ routine such as \texttt{polywind} in Numerical Recipes \mycitet{Pre07}, and holds regardless of the complexity of the occulter shape.  This approach also speeds up multiband calculations, as the geometric extent of the shadow determined this way is wavelength-independent.  Lateral errors in the occulter-telescope position may be included by adding a $\Delta \xi$ and $\Delta \eta$ to all points at the telescope plane, and that should be done first.  Note that in the case of an off-axis source, the geometric shadow is shifted laterally as well; in this case, the winding number should be calculated around $(\xi_0 - z \sin{\psi_1} \cos{\psi_2}, \eta_0 - z \sin{\psi_1} \sin{\psi_2})$.  We then have:
\begin{align} \label{Ug}
U^{(G)}(\xi_0, \eta_0, z) &= A \exp{(i k z \cos{\psi_1})} \exp{[i k (\xi_0 \cos{\psi_2} + \eta_0 \sin{\psi_2}) \sin{\psi_1}]}, \\
& \qquad \mathrm{for\;points\;outside\;the\;shadow} \nonumber \\
 & = 0 \; \mathrm{otherwise.} \nonumber
\end{align}
In some cases---for example, when the telescope aperture remains close to the center of the occulter and the perturbations to the ideal shape of the occulter are small---the winding number may be able to be determined independently.  This determination should be done if possible, as even an efficient algorithm can increase runtime significantly for boundaries containing a large number of points.

The boundary field can be evaluated efficiently with midpoint-rule quadrature running around the edge.  First, for each of the $N$ line segments running counterclockwise around the occulter edge, we derive the midpoint of line segment $j$ going from $(x_{j}, y_{j}, 0)$ to $(x_{j+1}, y_{j+1}, 0)$:
\begin{equation} \label{xmym}
    (x^{(m)}_{j}, y^{(m)}_{j}, 0) = ([x_{j} + x_{j+1}]/2, [y_{j} + y_{j+1}]/2, 0),
\end{equation}
and the vector pointing along the segment
\begin{equation} \label{xlyl}
    (x^{(\ell)}_{j}, y^{(\ell)}_{j}, 0)_j = (x_{j+1}-x_{j}, y_{j+1}-y_{j}, 0).
\end{equation}
with $(x_{j+1}, y_{j+1})$ equal to $(x_{1}, y_{1})$ when $j=N$.  This derivation need only be done once at the beginning, as it holds for all $M$ points and $L$ wavelengths.  (We note that, while the midpoint rule has been used for the quadrature, this is certainly not a requirement, and higher order methods could be used; \myeq{fullW} will still hold.  Using a higher-order method may come at the cost of runtime.)

For each point in the downstream field, we calculate $f$, $g$, and $h$ for all $N$ segments following \myeq{fgh}:
\begin{align}
f_j &= z \sin{\psi_1} + (x^{(m)}_{j} - \xi_0)\cos{\psi_2} \cos{\psi_1} + (y^{(m)}_{j} - \eta_0) \sin{\psi_2} \cos{\psi_1}, \label{vecfgh} \\
g_j &= -(x^{(m)}_{j} - \xi_0) \sin{\psi_2} + (y^{(m)}_{j} - \eta_0) \cos{\psi_2}, \\
h_j &= z \cos{\psi_1} - (x^{(m)}_{j} - \xi_0)\cos{\psi_2} \sin{\psi_1} - (y^{(m)}_{j} - \eta_0) \sin{\psi_2} \sin{\psi_1},
\end{align}
and we then need only two intermediate terms:
\begin{align}
T_{1j} &= f_j^2 + g_j^2,\\
T_{2j} &= \mathbf{v}(f_j,g_j) \cdot \mathbf{\ell} \intd{\ell} \nonumber \\
    &= x^{(\ell)}_{j} (z \sin{\psi_1} \sin{\psi_2} + (y^{(m)}_{j} - \eta_0) \cos{\psi_1}) -
    y^{(\ell)}_{j} (z \sin{\psi_1} \cos{\psi_2} + (x^{(m)}_{j} - \xi_0) \cos{\psi_1}), \label{terms}
\end{align}
to give us:
\begin{align} \label{Ubsum}
U^{(B)}(\xi_0, \eta_0, z) =& \frac{1}{2\pi} A \exp{(i k z \cos{\psi_1})} \exp{[i k (\xi_0 \cos{\psi_2} + \eta_0 \sin{\psi_2}) \sin{\psi_1}]} \nonumber \\
&\times \sum_{j = 1}^{N} \frac{\exp{\left(i k \frac{T_{1j}}{2 h_j}\right)}}{T_{1j}} T_{2j}.
\end{align}
Note that all of the steps prior to \myeq{Ubsum} are wavelength-independent; different values of $k$ may be iterated over in the final step.  It is advisable to keep terms of order $kz$ (e.g. $kz \cos{\psi_1}$ or $kh$) separate when evaluating to maintain numerical precision.  A typical implementation might take the form shown in Algorithm \ref{alg}.  The for-loops over $j$ in particular are well-suited to vectorization where this is supported.
\begin{algorithm}[ht]\label{alg}
\DontPrintSemicolon
\For{$j = 1$ \KwTo $N$ (all occulter edge points)}{
    Find $x^{(m)}_{j}, y^{(m)}_{j}, x^{(\ell)}_{j}, y^{(\ell)}_{j}$ using \myeq{xmym} and \myeq{xlyl}\;
}
\For{$q = 1$ \KwTo $M$ (all points at telescope aperture)}{
    Apply lateral error to point $q$, if any\;
    Get winding number around point $q$ with \texttt{polywind}\;
    \For{$j = 1$ \KwTo $N$ (all occulter edge points)}{
    Find $f_j, g_j, h_j, T_{1j}, T_{2j}$ using \myeq{vecfgh} through \myeq{terms}\;
    }
    \For{$s = 1$ \KwTo $L$ (all wavelengths)}{
    Find $U^{(B)}$ by summing over $h_j, T_{1j}, T_{2j}$ in \myeq{Ubsum}\;
    Find $U^{(G)}$ using the winding number and \myeq{Ug}\;
    Find $U = U^{(G)} - U^{(B)}$ for point $q$ and wavelength $s$\;
    }
}
\caption{Pseudocode representation of BDW calculation}
\end{algorithm}

\subsection{Computational results} \label{subsec:comp}

In the standard formulation of the Dubra-Ferrari integral (DF), points outside the geometric shadow are calculated with a different set of integrals which require searches for nearest- and further-neighbor points along the boundary for each point at the telescope aperture.  To simplify direct comparison between the Dubra-Ferrari approach and the boundary-diffraction-wave approach for this paper, we choose to select a region which lies entirely in the geometric shadow of the occulter.  This selection also fixes the winding number to $1$ at all points under consideration, letting us bypass an explicit calculation.  This choice is not an unreasonable assumption; the telescope will generally be entirely in the geometric shadow for a properly aligned occulter-telescope system.

Both approaches have similar formulations, with the primary difference coming in the fact that the Dubra-Ferrari integral uses angular coordinates $\theta$ centered about $(\xi_0, \eta_0)$, and so the angular distance between occulter edge points must be rederived for each point at the telescope aperture.  This calculation turns out to be one of the major sources of overhead during the DF calculation.

\begin{figure}[ht]
\begin{center}
\includegraphics[width=0.75\columnwidth]{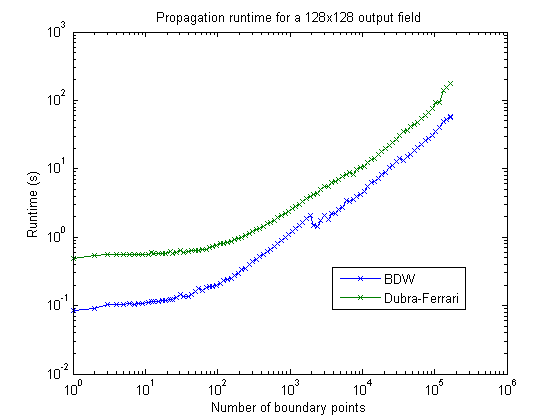}
\end{center}
\caption{A comparison of the time to run the two propagation algorithms, as a function of number of points along the edge of the entrance aperture.}\label{fig:run}
\end{figure}

\begin{figure}[ht]
\begin{center}
\includegraphics[width=0.75\columnwidth]{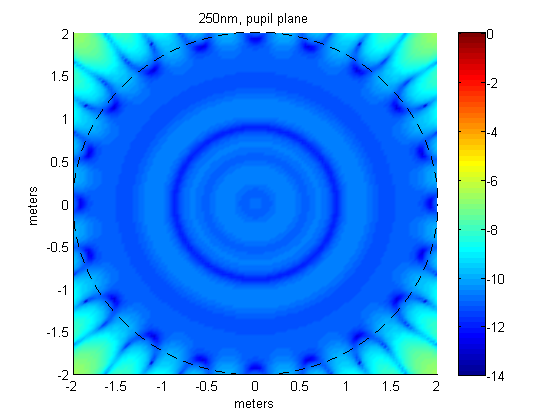}
\end{center}
\caption{Intensity at the plane of the telescope aperture following the occulter, plotted on a base-10 logarithmic scale.  The telescope aperture is shown by a dashed circle.}\label{fig:int}
\end{figure}

To compare the runtimes of the Dubra-Ferrari and BDW implementations, we consider a closed occulter boundary specified by $164000$ individual points, corresponding to the occulter shown in \myfig{fig:binocc}.  We select subsets of these points, and run Matlab implementations of both algorithms on this subset. \myfig{fig:run} shows the time to implement the propagation as a function of number of boundary points; the BDW algorithm remains 2-3$\times$ faster over a wide range of polygon sizes.  \myfig{fig:int} shows the intensity profile in a region around the telescope aperture, plotted on a base-10 logarithmic scale.  The plots of this intensity profile created with Dubra-Ferrari and BDW are not visually distinguishable; within the telescope aperture, the maximum difference in intensity is $4.8\times10^{-15}$.

\section{Conclusion}

We present a new approach to modeling fields following planet-finding occulters, based on the boundary-diffraction-wave formulation laid out by Miyamoto and Wolf.  It has proved to be simple to modify the shape and recalculate with this approach to include various types of occulter errors, as well as fast to evaluate, improving on the current best method by 2-3$\times$ with no loss in precision.  An efficient implementation has been laid out, and we hope it proves useful to others in the field.

\section*{Acknowledgments}

This work was carried out as a part of the Imaging Performance Study funded by NASA's Exoplanet Exploration Program; the research was performed at the Jet Propulsion Laboratory, California Institute of Technology, under a contract with the National Aeronautics and Space Administration.


\begin{thebibliography}{10}

\bibitem{Kas11}
N.~J. {Kasdin}, D.~N. {Spergel}, R.~J. {Vanderbei}, D.~{Lisman}, S.~{Shaklan},
  M.~{Thomson}, P.~{Walkemeyer}, V.~{Bach}, E.~{Oakes}, E.~{Cady}, S.~{Martin},
  L.~{Marchen}, B.~{Macintosh}, R.~E. {Rudd}, J.~{Mikula}, and D.~{Lynch},
  ``Advancing technology for starlight suppression via an external occulter,''
    Proc. SPIE 8151, 81510J (2011).

\bibitem{Sha10}
S.~B. {Shaklan}, M.~C. {Noecker}, T.~{Glassman}, A.~S. {Lo}, P.~J.
  {Dumont}, N.~J. {Kasdin}, E.~J. {Cady}, R.~{Vanderbei}, and P.~R.
  {Lawson}, ``Error budgeting and tolerancing of starshades for exoplanet
  detection,''
  Proc. SPIE 7731, 77312G (2010).

\bibitem{Gla10}
T.~{Glassman}, A.~{Johnson}, A.~{Lo}, D.~{Dailey}, H.~{Shelton}, and
  J.~{Vogrin}, ``Error analysis on the NWO starshade,''
    Proc. SPIE 7731, 773150 (2010).

\bibitem{Van07}
R.~J. Vanderbei, E.~J. Cady, and N.~J. Kasdin, ``Optimal occulter design for finding extrasolar planets,''
\newblock Astrophys. J. \textbf{665}, 794--798 (2007).

\bibitem{Cad08}
E.~{Cady}, L.~{Pueyo}, R.~{Soummer}, and N.~J. {Kasdin}, ``Performance of hybrid occulters using apodized pupil Lyot
  coronagraphy,''
Proc. SPIE 7010, 70101X (2008).

\bibitem{Cop00}
C.~J. Copi and G.~D. Starkman, ``The Big Occulting Steerable Satellite [{B}{O}{S}{S}],''
\newblock  Astrophys. J. \textbf{532}, 581--592 (2000).

\bibitem{Cas06}
W.~Cash, ``Detection of earth-like planets around nearby stars using a
  petal-shaped occulter,''
\newblock Nature \textbf{442}, 51--53 (2006).

\bibitem{Goo96}
J.~W. Goodman, {\em Introduction to Fourier Optics} (McGraw-Hill, 1996).

\bibitem{Bor99}
M.~Born and E.~Wolf, {\em Principles of Optics} (Cambridge University Press, 1999).

\bibitem{Sou07}
R.~Soummer, L.~Pueyo, A.~Sivaramakrishnan, and R.~J. Vanderbei, ``Fast computation of Lyot-style coronagraph propagation,''
\newblock Opt. Express \textbf{15}, 24, 15935--15951, (2007).

\bibitem{Cad10b}
E.~{Cady}, ``Design, tolerancing, and experimental verification of occulters
  for finding extrasolar planets,''
\newblock PhD thesis, Princeton University, 2010.

\bibitem{Dub99}
A.~{Dubra} and J.~A. {Ferrari}, ``Diffracted field by an arbitrary aperture,''
\newblock Am. J. of Phys. \textbf{67}, 87--92 (1999).

\bibitem{Dum09}
P.~Dumont, S.~Shaklan, E.~Cady, J.~Kasdin, and R.~Vanderbei, ``Analysis of external occulters in the presence of defects,''
Proc. SPIE 7440, 744008 (2009).

\bibitem{Miy62}
K.~{Miyamoto} and E.~{Wolf}, ``Generalization of the Maggi-Rubinowicz theory of the boundary
  diffraction wave part I,''
\newblock J. Opt. Soc. Am (1917-1983) \textbf{52}, 615 (1962).

\bibitem{Miy62a}
K.~{Miyamoto} and E.~{Wolf}, ``Generalization of the Maggi-Rubinowicz theory of the boundary
  diffraction wave part II,''
\newblock J. Opt. Soc. Am (1917-1983) \textbf{52}, 626 (1962).

\bibitem{Pre07}
W.~H. {Press}, S.~A. {Teukolsky}, W.~T. {Vetterling}, and B.~P. {Flannery}, {\em {Numerical Recipes. The Art of Scientific Computing}} (Cambridge University Press, 2007).

\end{thebibliography}
\end{document}